\documentstyle[titlepage,12pt]{article}
\newcommand{\beq}{\begin{equation}}
\newcommand{\eeq}{\end{equation}}

\newcommand{\eq}[1]{eq.(\ref{#1})}
\title {
\begin{flushright} PSU/TH/155
\end{flushright}
\bigskip\bigskip
CORRECTIONS OF ORDER $\alpha^2(Z\alpha)^5$ TO HYPERFINE SPLITTING
AND LAMB SHIFT}

\medskip

\author {Michael I. Eides\thanks{E-mail address:
eides@lnpi.spb.su, eides@phys.psu.edu}\\
Department of Physics, Pennsylvania State University,\\
University Park, Pennsylvania 16802, USA\thanks{Temporary address}\medskip\\
and\\
Petersburg Nuclear Physics Institute,\\
Gatchina, St.Petersburg 188350, Russia\thanks{Permanent address}\medskip\\
and
\and Valery A. Shelyuto\\
D. I.  Mendeleev Institute of Metrology, \\ St.Petersburg
198005, Russia}
\date{January, 1995}

\begin{document}

\maketitle

\begin{abstract}
Corrections to hyperfine splitting and Lamb shift of order
$\alpha^2(Z\alpha)^5$ induced by the diagrams with radiative photon
insertions in the electron line are calculated in the Fried-Yennie gauge.
These contributions are as large as
$-7.725(3)\alpha^2(Z\alpha)^5/(\pi n^3)(m_r/m)^3m$ and
$-0.6711(7)\alpha^2(Z\alpha)/(\pi n^3)E_F$ for the Lamb shift and
hyperfine splitting, respectively. Phenomenological implications of these
results are discussed with special emphasis on the accuracy of the
theoretical predictions for the Lamb shift and experimental determination of
the Rydberg constant. New more precise value of the Rydberg constant is
obtained on the basis of the improved theory and experimental data.
\end{abstract}

A steady and rapid progress in the spectroscopic measurements in recent
years led to a dramatic increase of accuracy in the measurements of the
Rydberg constant \cite{and,nez}, ground state $1S$ Lamb shift in hydrogen
and deuterium \cite{nez,weitz,bosh}, classic $2S-2P$ Lamb shift in
hydrogen \cite{lp,sok,hp}, and of the muonium hyperfine splitting in the
ground state \cite{mariam,hughes} (see Table 1).

These spectacular experimental achievements constitute a serious challenge
to the theory and intensive theoretical efforts are necessary to match this
experimental accuracy.

Theoretical work on the high order corrections to hyperfine splitting (HFS)
and Lamb shift concentrated recently on calculation of nonrecoil
contributions of order $\alpha^2(Z\alpha)^5$. Their magnitude may run up to
several kilohertz for HFS in the ground state of muonium,  to several
tens of kilohertz for $n=2$ Lamb shift in hydrogen and may be as large as
hundreds of kilohertz for the ground state Lamb shift in hydrogen.
Contributions of such order of magnitude are clearly crucial for comparison
of the current and pending experimental results with the theory.

As was shown in \cite{eks1} for hyperfine splitting and in \cite{ego} for the
Lamb shift there are six gauge invariant sets of diagrams (see Fig.1), which
produce corrections of order $\alpha^2(Z\alpha)^5$. All these diagrams may
be obtained  from the  skeleton diagram, which contains two external photons
attached to the electron line, with the help of different radiative
insertions. All contributions  induced by the diagrams in Figs.$1a-1e$,
containing closed electron loops, were obtained recently in papers
\cite{eks1,eks2,eks5,kn} for the case of hyperfine splitting and in papers
\cite{ego,eg,eg4,eg5,pach1} for the case of the Lamb shift. These
theoretical results are now firmly established since all these corrections
were calculated independently by two different groups and the results of
these calculations are in excellent agreement.

We report below on the results of our calculation of the contributions of
order $\alpha^2(Z\alpha)^5$ to HFS and Lamb shift induced by the last
gauge invariant set of diagrams in Fig.$1f$. This set includes nineteen
topologically different diagrams \cite{eksl1} presented in Fig.2.
The simplest way to describe these graphs is to realize that they were
obtained from the three graphs for the two-loop electron self-energy by
insertion of two external photons in all possible ways. Really, graphs
$2a-2c$ are obtained from the two-loop reducible electron self-energy
diagram, graphs $2d-2k$ are the result of all possible insertions of two
external photons in the rainbow self-energy diagram, and diagrams $2l-2s$
are connected with the overlapping two-loop self-energy graph.
We have already calculated contributions induced by the diagrams in
Figs.$2a-2h$ and Fig.$2l$ earlier \cite{eksl1,eksl2}. Results of the
calculation of the contributions produced by the remaining diagrams in Fig.2
are presented below.

Let us start with a brief description of the main features of our approach
to calculations.  As was shown in \cite{ann1}  for HFS and in \cite{ego} for
the Lamb shift, contributions to the energy splittings are given by the
matrix elements of the diagrams in Fig.2  calculated between free electron
spinors with all external electron lines on the mass shell, projected
on the respective spin states and multiplied by the square at the origin of
the Schr\"{o}dinger-Coulomb wave function.

Actual calculation of the matrix elements is impeded by the ultraviolet
and infrared divergences. Infrared problems are as usual more difficult to
deal with than the ultraviolet ones. It is easy to realize that in the
standard Feynman gauge all diagrams in Fig.2 are infrared divergent and one
has to introduce the radiative photon mass to regularize this divergence.
Sure, the final result for the sum of all contributions induced by the
diagrams in Fig.2 is infrared finite and should admit a smooth limit for
the vanishing photon mass. However, numerical recipes used in calculations
of the contributions to the energy shifts make it impossible to check
analytically independence of the results on the photon mass and one has to
rely on the extrapolation in the infrared photon mass. Of course, such
approach is still feasible, but we have preferred to use the gauge
invariance of the sum of diagrams in Fig.2 and to perform all calculations
in the Fried-Yennie (FY) gauge \cite{fy} for the radiative photons. All
diagrams are infrared finite in this remarkable gauge and one may perform
the on-mass-shell renormalization without introduction of the infrared
photon mass (see, e.g., \cite{ann1}) avoiding thus the problem of
extrapolation to the vanishing photon mass. Of course, infrared finiteness
in the FY gauge is not given for free, and one has to pay special attention
to the infrared behavior of the integrand functions and to perform
cancellation of spurious infrared divergences with the help of integration
by parts over the Feynman parameters prior to momentum integration.

Calculation of the contributions to the energy splittings starts with
putting down the universal infrared diverging skeleton integrals
corresponding to the electron line with two external photons. Each
contribution of order $\alpha^{2}(Z\alpha)^{5}$ arises from radiative
insertions in the skeleton graph. Corrections to hyperfine splitting and
Lamb shift, produced by the diagrams in Figs.1 and 2 are given  by  the
expressions (see, e.g.  \cite{ann1,ego})

\beq                                  \label{hfs}
\Delta E_{HFS} = 8~\frac{Z\alpha}{\pi n^3}(\frac{\alpha}{\pi})^2\:E_F
\int_0^\infty\frac{dk}{{k}^2}\:L_{HFS}(k)\:,
\eeq

and

\beq                    \label{lamb}
\Delta E_L =-16~\frac{(Z\alpha)^5}{\pi n^3 }\left(\frac{\alpha}{\pi}\right)^2
\:\left(\frac{m_r}{m}\right)^3\:m\int_0^\infty\frac{dk}
{k^4}\:L_{L}(k)\:,
\eeq

where $k$ is the magnitude of the three-dimensional momentum of the external
photons measured in the electron mass units, $m_r={m}/(1+{m}/{M})$ is the
reduced mass of the electron-muon (or electron-proton) system and $E_F$ is
the Fermi energy of hyperfine splitting. The functions $L(k)$ are
connected with the numerator structure and spin projection of each
particular graph and describe radiative corrections to the skeleton diagram.
They are normalized on the skeleton numerator contributions.

It should be mentioned that some of the diagrams under consideration
also contain contributions of the previous order in $Z\alpha$. Physical
nature of these contributions is especially transparent in the case of HFS.
They correspond to anomalous magnetic moment, their true order in $Z\alpha$
is lower than their apparent order and they should be subtracted from the
electron factor prior to calculation of the contributions to HFS. Analogous
situation holds also in the case of the Lamb shift. The only difference is
that this time not only the Pauli formfactor but also the slope
of the Dirac formfactor of the electron is capable to produce lower order
contribution to the splitting of the energy levels (see, e. g. \cite{ego},
\cite{eg} and \cite{eg4}). Technically cases of lower order contributions
both to HFS and to the Lamb shift are quite similar.  Lower order terms are
produced by the constant terms in the low-frequency asymptotic expansion of
the electron factor in the case of the hyperfine splitting and by the terms
proportional to the exchanged momentum squared in the low-frequency
asymptotic expansion of the electron factor in the case of the Lamb shift.

These lower order contributions are connected with integration
over external photon momenta of characteristic atomic order $mZ\alpha$ and
the approximation based on the skeleton integrals in \eq{hfs} and \eq{lamb}
is unadequate for their calculation. In the skeleton integral approach these
previous order contributions emerge as the infrared divergences induced by
the low-frequency terms in the electron factors. We subtract leading
low-frequency terms in the low-frequency asymptotic expansions of the
electron factors, when necessary, and thus get rid of the previous order
contributions.

The results of our calculations of the contributions to  HFS and Lamb
produced by different diagrams in the FY gauge are presented in Table 2
\footnote {Detailed account of our calculations will be presented in a
separate publication.}. For the total correction of order
$\alpha^2(Z\alpha)^5$ to the HFS and the Lamb shift produced by all diagrams
in Fig.2 we obtain

\beq                    \label{hfsf}
\Delta E^{(1f)}_{HFS}=-0.~6711~(7)~\frac{\alpha^2(Z\alpha)}{\pi n^3}~E_F,
\eeq

and

\beq                     \label{lambf}
\Delta E^{(1f)}_L=-7.~725~(3)~\frac{\alpha^2(Z\alpha)^5}{\pi n^3}
{}~(\frac{m_r}{m})^3~m.
\eeq

While this work was in progress two other papers were published where the
contributions of the diagrams in Fig.2 to HFS \cite{kn} and the Lamb shift
\cite{pach2} were calculated. The authors of these works used a completely
different approach to calculations, in particular, they worked in the
Feynman gauge and, hence, all contributions of individual diagrams in
these works are infrared divergent.  The numbers cited in \cite{kn,pach2}
are obtained with the help of extrapolation of the presumably infrared
finite sum of all contributions in Fig.2 to the vanishing infrared photon
mass. Despite the great differences in the approaches used in the present
work and in \cite{kn,pach2} numerical factors in \eq{hfsf} and \eq{lambf}
are compatible with $-0.63(4)$ in \cite{kn} and with $-7.61(16)$ in
\cite{pach2}, respectively. Our numbers are about two orders of magnitude
more precise and further improvement of accuracy may be achieved. The
reason for this increased accuracy is the use of the FY gauge, where one can
avoid the extrapolation in the photon mass. The price we had paid for this
advantage is the more tiresome analytic work needed to cancel all would be
infrared divergences before integration.

Numerically the correction to muonium HFS in the ground state produced by
the diagrams in Fig.2 is equal to

\beq
\Delta E^{(1f)}_{HFS} =-0.~3701~(4)~\mbox{kHz}.
\eeq

and the total contribution of order $\alpha^2(Z\alpha)E_F$ is given by

\beq
\Delta E^{(1a-1f)}_{HFS} =0.~4264~(4)~\mbox{kHz}.
\eeq

Taking into account other theoretical contributions to HFS and especially
some small contributions obtained recently (see, e.g., reviews in
\cite{kn,eid}) and using for calculation the value of the fine structure
constant as obtained in \cite{cage} one may obtain the theoretical value
for the muonium HFS in the ground state

\beq
\Delta E_{HFS} =4~463~302.~55~(0.18)~(0.18)~(1.33)~\mbox{kHz},
\eeq

where the first error in parenthesis reflects the uncertainty of the fine
structure constant itself and the second is induced by the uncertainty of
the contribution of order $\alpha(Z\alpha)^2E_F$. The third, and by
far the largest contribution to the error in the theoretical value of HFS is
defined by the experimental error in measuring electron-muon mass ratio
$m/M$.

The agreement between theory and experiment is excellent. We will not  dwell
on the HFS problem any more here since the phenomenological situation and
the influence of the result in \eq{hfsf} on the value of the electron-muon
mass ratio and the fine structure constant was discussed in great detail
recently \cite{kn,eid}.

The case of Lamb shift deserves more comments. Numerically the corrections
to the $1S$ and $2S$ Lamb shifts produced by the diagrams in Fig.2 are
equal to

\beq                        \label{lambfn}
\Delta E^{(1f)}_{L}(1S) =-334.~2~(1)~\mbox{kHz},
\eeq
\[
\Delta E^{(1f)}_{L}(2S) =-41.~78~(2)~\mbox{kHz},
\]

while respective total contributions of order $\alpha^2(Z\alpha)^5m$ are
given by

\beq
\Delta E^{(1a-1f)}_{L}(1S) =-296.~9~(1)~\mbox{kHz},
\eeq
\[
\Delta E^{(1a-1f)}_{L}(2S) =-37.~12~(2)~\mbox{kHz}.
\]

Let us discuss the sign and the scale of the correction in \eq{lambfn}. The
sign may be determined by considering the electron factor, as defined in
\eq{lamb}.  The low-frequency asymptotic behavior of the electron factor
is described by the expresssion (see, e.g., \cite{eg4})

\beq                             \label{ff}
L( k)\approx (-2F_1'(0)-\frac{1}{2}F_2(0))~k^2=-0.7046225~ k^2,
\eeq

where $F_1(k^2)$ and $F_2(k^2)$ are the two-loop contributions  to
the ordinary Dirac and Pauli form factors of the electron
(contribution of the graphs with vacuum polarization insertions in the
photon line is omitted in \eq{ff}), respectively. We use in \eq{ff}
the well-known values for the slope of the Dirac form factor and of the
Pauli form factor at zero \cite{ab,bmr}. As was explained above one has to
subtract from the electron factor this leading low-frequency term which
produces contribution to the Lamb  shift of previous order in $Z\alpha$.

It is well known from the general principles that the unsubtracted electron
factor has at most logarithmic behavior at infinity. Hence, the high
momentum behavior of the subtracted electron factor is completely defined by
the subtraction term in \eq{ff}. Then it is clear from \eq{lamb}, where the
subtracted electron factor plays the role of the integrand, that the
contribution to the Lamb shift induced by the graphs in Fig.2 has the
negative sign. One may even make an estimate of this contribution from the
known asymptotic behavior of the integrand but we choose a less technical
path in discussion of the magnitude of this contribution.

It may seem at first sight that the magnitude of the corrections induced by
the diagrams in Fig.$1f$ as presented in \eq{lambfn} are too large. We would
like to emphasize that, quite opposite, this correction has exactly the
scale one had to envisage before calculations. Let us discuss this point in
slightly more detail. It is helpful to recollect that the main
contribution to the Lamb shift is a radiative correction itself and so it is
misleading to normalize all contributions to the Lamb shift with the help of
this leading order contribution. In this respect the case of the Lamb shift
differs drastically from the case of HFS, where the leading Fermi
contribution is not the radiative correction but the classic effect of the
interaction of two magnetic moments and sets the natural scale for all
radiative corrections.  Main contribution to the Lamb shift has the form
$4m(Z\alpha)^4/n^3\times\mbox{\sl slope of the Dirac formfactor}$, where the
slope is roughly speaking $\alpha/\pi\cdot(1/3)\cdot\ln(Z\alpha)^{-2}$.  The
skeleton factor which sets the scale for the different contributions to the
Lamb shift is $4m(Z\alpha)^4/n^3$ and to make an estimate of any
correction to the Lamb shift one has to extract this skeleton factor. All
other entries in the leading order contribution to the Lamb shift are
produced by the radiative correction, and it is necessary to take into
account that the number which should be of order one, as predicts the common
wisdom for radiative corrections, is the factor $1/3$ before the logarithm
which remains after extraction of the factor $\alpha/\pi$, characteristic
for the one-loop radiative corrections.  The other subtlety to be taken into
account in estimating the orders of magnitude of different corrections is
that, unlike the case of radiative corrections to the scattering amplitudes,
in the bound state problem not every factor $\alpha$ is accompanied by an
extra factor $\pi$ in the denominator. This is a well known feature of the
Coulomb problem.

Let us consider as an exercise in the art of making educated estimates
the correction of order $\alpha(Z\alpha)^5m$ calculated analytically long
time ago \cite{kks,bbf}.  According to the considerations above the scale of
this correction should be set by the factor $4\alpha(Z\alpha)^5/n^3m$ and
the only problem of the theory is to calculate the number of order one
before this factor. Analytic calculation \cite{kks,bbf} produces this
factor in the form $1+11/128-1/2\log 2\approx 0.739$ in excellent agreement
with our qualitative considerations. Now it is easy to realize that the
natural scale for the correction of order $\alpha^2(Z\alpha)^5$ is set by
the factor $4\alpha^2(Z\alpha)^5/(\pi n^3)m$. The coefficient before this
factor obtained above and in \cite{pach2} is about $-1.9$ and there is
nothing unusual in its magnitude for a numerical factor corresponding to a
radiative correction.

Consider now current status of the Lamb shift theory.
Theoretical predictions presented below are
obtained with the help of the expressions for the Lamb shift
contributions as collected in the reviews \cite{sy,gr}, amended, besides
corrections obtained above and in \cite{pach2}, with some other recent
results presented in the Table 3. Note that the correction of order
$\alpha^2(Z\alpha)^6$ in this Table is again of reasonable magnitude since
its scale is set by the factor $4\alpha^2(Z\alpha)^6/(\pi^2 n^3)m$ as one
may easily check with the help of the arguments used above in the discussion
of the contribution of order $\alpha^2(Z\alpha)^5$. On the background
of this factor numerical factor $2/27$ before the logarithm cube is quite
moderate.

Last line in Table 3 contains a new recoil correction corresponding to the
insertions in the Coulomb photon of the muon or hadron vacuum polarization
operators\footnote {K. Pachucki and S.
Karshenboim are also considering these contributions (private communication
from S.  Karhenboim).}. Respective contribution for the muon insertion
contains an evident extra electron-muon mass ratio squared suppression
factor relative to the leading vacuum polarization contribution. We
estimated hadron contribution approximating the spectral function below 1
GeV according to the vector dominance model and above 1 GeV we simply used
the asymptotic quark value for the spectral function\footnote{The derivation
of this result will be published elsewhere.}.

We use in calculation of the theoretical values for the Lamb shift new
values for the self-energy contributions to the coefficient $C_{60}$ for the
$1S$- and $2S$-states \cite{pach3}, and to the function $G_{SE}$ for the
$2P_{1/2}$-state \cite{kar2} and for the $4S_{1/2}$-state \cite{mohr}. We
also use new values \cite{mohr} $G_{VP}(1S_{1/2})=-0.6187$,
$G_{VP}(2S_{1/2})=-0.8089$, $G_{VP}(2P_{1/2})=-0.0640$ and
$G_{VP}(4S_{1/2})=-0.8066$ for the Uehling part of the vacuum polarization
contribution. These numbers are the sum of contributions of order
$\alpha(Z\alpha)^6$ and of additional terms of higher order in $Z\alpha$.

{}From the theoretical point of view the accuracy of calculations is limited
by the magnitude of the yet uncalculated contributions to the Lamb shift.
First, there are pure recoil contributions of order $(Z\alpha)^6(m/M)m$
which may be as large as $1.3$ kHz for the $1S$-state and $0.16$ kHz for the
$2S$-state. As far as we know, such terms were never discussed in the
literature \cite{sy}. It is claimed in \cite{pach4} that there are no such
contributions besides recoil corrections obtained in \cite{grp}, but,
unfortunately, no proof of this statement was published. We will accept that
such corrections are absent, especially taking into account that according
to the estimates above they are in any case too small to influence really
the comparison of the theory with the experimental results.

Unknown correction of order $\alpha^3(Z\alpha)^4$ is induced
by the three-loop slope of the Dirac formfactor of the electron and by the
three-loop electron vacuum polarization. Natural scale for this correction
is set by the factor $4\alpha^3(Z\alpha)^4/(\pi^3n^3)m$ and we envisage the
contributions about $17$ kHz for the $1S$-state and $2$ kHz for the
$2S$-state.

Next come uncalculated corrections of order $\alpha^2(Z\alpha)^6$.
Contribution of this order is a polynomial in $\ln(Z\alpha)^{-2}$,
starting with log cube. The factor before log cube was calculated in
\cite{kar2} and the contribution of the log squared terms to the
difference $E_L(1S)-8E_L(2S)$ was obtained in \cite{kar1}. However, the
calculation of respective contributions to the separate energy levels is
still missing. In this conditions it is fair to take the log
squared contribution to the interval $E_L(1S)-8E_L(2S)$ as an estimate of
the scale of all yet uncalculated corrections of this order. We thus assume
that uncertainties induced by the yet uncalculated contributions of order
$\alpha^2(Z\alpha)^6$ constitute $15$ kHz and $2$ kHz for the $1S$- and
$2S$-states, respectively.

The scale of the self-energy correction of order $\alpha(Z\alpha)^7$ is set
by the factor  $4\alpha(Z\alpha)^7/n^3$. This contribution is a linear
polynomial in $\ln(Z\alpha)^{-2}$. We are aware about two recent attempts to
make an estimate of this contribution \cite{drake,kar1}, but, unfortunately,
its final magnitude seems to be still unavailable\footnote{P. Mohr is now
working on the extraction of this correction from his respective high
$Z\alpha$ results (Private communication from P. Mohr).}. Relying on the
scale factor above and the estimates in \cite{drake,kar1} we assume that the
corrections of order $\alpha(Z\alpha)^7$ are as large as $17$ kHz and $2$
kHz to the $1S$- and $2S$-states, respectively.

All other theoretical contributions to the Lamb shift are
smaller than those just discussed. Hence, we assume that the theoretical
uncertainty of the expression for the Lamb shift is about $28$ kHz for the
$1S$-state and about $4$ kHz for the $2S$-state.

The other limit on the accuracy of the theoretical calculation of the Lamb
shift is put by the accuracy of the measurements of the proton
rms charge radius. As is well known there are two contradictory experimental
results for this radius \cite{dhr,ssbw} and at least one of these
experimental results should be in error. The accuracy of the proton rms
charge radius claimed by the authors of \cite{dhr,ssbw} produces uncertainty
about $32$ kHz for the $1S$-state and about $4$ kHz for the $2S$-state.

Let us compare theoretical and experimental data for the classic
$2S_{1/2}-2P_{1/2}$ Lamb shift. The most precise experimental data as well
as the results of our theoretical calculations are presented in Table 4.
Theoretical results for the energy shifts in Table 4 contain errors in
the parenthesis where the first error is determined by the yet uncalculated
contributions to the Lamb shift, discussed above, and the second reflects
the experimental uncertainty in the measurement of the proton rms charge
radius. We have used experimental result \cite{sok} taking into account
recent theoretical correction discovered in \cite{kar1}.
Note, however, that this correction does not effect any of our conclusions.
There are two immediate conclusions of the data in Table 4. First, as
already mentioned in \cite{pach2}, the results of the proton rms radius
measurement in \cite{dhr} should be in error since respective value of the
proton charge radius is clearly inconsistent with all results of the Lamb
shift measurements. Second, we have to reject either the result of the most
precise measurement of the $2S_{1/2}-2P_{1/2}$ splitting, or the experimental
value
of the proton charge radius as measured in \cite{ssbw} since the Lamb
shift value in \cite{sok} contradicts theoretical value calculated
employing the rms radius in \cite{ssbw} by more than five standard
deviations. Results of two other measurements of the classic Lamb shift are
compatible with the theory, so we will below accept the value of the
proton charge radius as obtained in \cite{ssbw}. We will return to the
numbers in three last lines in Table 4 below.

We do not include theoretical predictions for the deuterium Lamb
shift in Table 4, since, taking into account current discrepancies in
determination of the deuteron charge radius and solid status of the Lamb
shift theory, it seems preferable to use the deuteron Lamb shift data for
extracting the value of this charge radius rather than for comparison of the
Lamb shift theory and experiment.

Next we turn to the discussion of the $1S$ Lamb shift. Unfortunately, its
extraction from the experimental data is less straightforward. The
experimentalists managed to separate measurement of the $1S$ Lamb shift from
the measurement of the Rydberg constant by comparing the frequencies
of two transitions with different main quantum numbers and excluding the
large levels separation depending on the Rydberg constant. In this manner
certain combinations of $1S$, $2S$ and of higher levels Lamb shifts are
experimentally obtained. It is pretty easy to compare these experimental
data in \cite{and,nez,weitz,bosh} with the theory above and after trivial
calculations we have found an excellent agreement between the theory and
experiment. We will not put down these results here.

Unbiased extraction of the $1S$ Lamb shift from the experimental data is
still a problem. It is impossible to avoid using to this end the
experimental value of the $2S$ Lamb shift, and the emerging value of the
$1S$ Lamb shift depends thus on the experimental result for the
$2S_{1/2}-2P_{1/2}$ splitting. Higher levels Lamb shifts, which also enter
the problem may be safely calculated with sufficient accuracy. The standard
approach accepted by all experimental groups consists in adopting one or the
other $2S_{1/2}-2P_{1/2}$ experimental result and extracting thus the value
of the $1S$ Lamb shift.  All values in Table 1 for the $1S$ Lamb shift are
obtained in this manner with the help of experimental values in \cite{lp} or
in \cite{hp} for the classic Lamb shift.  These values should be compared
with our theoretical prediction

\beq    \label{1sth}
\Delta E_L(1S)=8~172~729~(28)~(32)~\mbox{kHz},
\eeq

where again the first error is determined by the yet uncalculated
contributions to the Lamb shift and the second reflects the experimental
uncertainty in the measurement of the proton rms charge radius.

The results of all experiments mentioned in Table 1 are pretty
consistent and their agreement with the theoretical value in \eq{1sth} is
satisfactory but not spectacular. It is necessary to recollect at
this point that the "experimental values" in the Table depend on the
experimental value of the $2S_{1/2}-2P_{1/2}$ Lamb shift adopted in their
extraction and the change of the $1S$ Lamb shift value under transition from
one experimental $2S_{1/2}-2P_{1/2}$ Lamb shift value to another may be
significant.

It would be helpful to find a way to extract the value of the $1S$ Lamb
shift from the experimental data unambiguously without reference to the
$2S_{1/2}-2P_{1/2}$ experimental results, which, while being compatible with
the theory, seem to be somewhat larger than the theoretical prediction. A
natural way to reach this goal is to use the theoretical relation between
the $1S$ and $2S$ Lamb shifts. A good deal of theoretical contributions to
the Lamb shift scale as $n^3$ and, hence, vanish in the difference
$8E_L(2S)-E_L(1S)$. In particular, all main sources of the theoretical
uncertainty, namely, proton charge radius contribution and almost all yet
uncalculated corrections to the Lamb shift mentioned above vanish.
Significant contribution to this difference may be produced only by the term
of the form $\alpha^2(Z\alpha)^6\ln(Z\alpha)^{-2}$, which was calculated
recently \cite{kar1}. This term produces correction about $14$ kHz and the
accuracy of the difference under consideration is determined by the yet
uncalculated single log contribution of the same order. Such term would not
change the log squared term by more than fifty percent and, hence, the
uncertainty of the difference under consideration is about $7$ kHz.  Hence,
we obtain the relation

\beq
8E_L(2S)-E_L(1S)=\Delta,
\eeq

where $\Delta=187~234~(7)~\mbox{kHz}$.

Now one may obtain self-consistent values for the $1S$ Lamb shift directly
from the experimental data in Refs.\cite{weitz,bosh,nez} with the help of
the relations

\beq
E_L(1S)=\frac{8}{3}(F_{1S-2S}-4F_{2S-4S_{1/2}})
-\frac{32}{3}E_L(4S_{1/2})+\frac{5}{3}\Delta,
\eeq
\beq
E_L(1S)=\frac{8}{3}(F_{1S-2S}-4F_{2S-4P_{1/2}})
-\frac{32}{3}E_L(4P_{1/2})+\frac{5}{3}\Delta,
\eeq
\beq
E_L(1S)=\frac{40}{19}(F_{1S-2S}-\frac{16}{5}F_{2S-8D_{5/2}})
-\frac{128}{19}E_L(8D_{5/2})+\frac{21}{19}\Delta.
\eeq

Numerical results are presented in Table 5. These results have
somewhat larger errors than the respective results in Table 1, however, they
do not depend on the experimental value of the $2S_{1/2}-2P_{1/2}$ Lamb
shift and on the value of the proton charge radius. The accuracy of the
self-consistent numbers in Table 5 is mainly determined by the
accuracy of the frequency measurements in \cite{weitz,nez,bosh}.  Factor
4-5 reduction of the experimental errors would lead to a self-consistent
determination of the $1S$ Lamb shift with the same accuracy as for the
values cited in Table 1. One may even invert the usual approach to the
$2S_{1/2}-2P_{1/2}$ and $1S$ Lamb shift values and extract values of the
2S-2P Lamb shift from the respective self-consistent $1S$ values (see three
last lines in Table 4).  These values of the $2S_{1/2}-2P_{1/2}$ Lamb shift
are consistent with the results of the direct measurements of the
$2S_{1/2}-2P_{1/2}$ Lamb shift but have somewhat larger error bars.
However, self-consisitent values of the $2S_{1/2}-2P_{1/2}$ Lamb shift would
become quite competitive with the results of the direct measurements after
the 4-5 times reduction of the current experimental errors in the frequency
measurements would be achieved.

Reduction of the errors of the values of the $1S$ and $2S_{1/2}-2P_{1/2}$
Lamb shifts opens new ways to a more precise determination of the Rydberg
constant. We would like to mention two new directions in the determination
of the Rydberg constant value besides the one adopted now (see, e.g.,
\cite{and,nez}). First, one can use the self-consistent value of the $1S$
Lamb shift and respective $2S_{1/2}-2P_{1/2}$ Lamb shift to get the value of
the Rydberg constant. Today such approach leads to a loss of accuracy in
comparison with the current experimental value of the Rydberg constant (see
Table 6, where the first error in the self-consistent values of the
Rydberg constant is determined by the accuracy of the self-consistent
Lamb shift values and the second is determined by the accuracy of
the frequency measurement), but greater accuracy may be achieved in future.
Important advantage of such approach is that the value obtained in
this way is independent of the direct experimental results on
$2S_{1/2}-2P_{1/2}$ Lamb shift and of the value of the proton charge radius.
Second new approach is simply to reject the experimental data on the Lamb
shifts and to use for the determination of the Rydberg constant directly the
data on the frequencies of transitions between the levels with different
main quantum numbers. Such approach becomes feasible now since the
accuracy of the theoretical expression for such transitions is defined by
the theoretical error of the expression for the $1S$ (or $2S$) Lamb shift
which is about $28$ kHz (and is even smaller for the $2S$ Lamb shift) as
discussed above  and is thus smaller than the experimental error of the
frequency determination.  Respective values of the Rydberg constant are
again presented in Table 6, where the first error is determined by the
accuracy of the theoretical expression, the second is defined by the
experimental error of the frequency measurement, and the third one is
determined by the experimental error in the determination of the proton
charge radius. The values of the Rydberg constant in two last lines in Table
6 derived from independent experimental data \cite{nez,and} are pretty
consistent. These values are more accurate than the ones obtained by other
methods and are the most precise contemporary values of this constant.
Natural drawback of this approach is, of course, the dependence of the
obtained value of the Rydberg constant on the proton charge radius.

In conclusion we would like to emphasize that the high accuracy of the
Lamb shift theory opens new perspectives in determination of the
Rydberg constant and of the Lamb shift in the $1S$- and $2S$-states. Four
directions of experimental investigations, namely, more precise measurement
of the transitions between levels with different main quantum numbers, more
precise measurement of the $1S$ and $2S$ Lamb shifts, and direct measurement
of the proton charge radius seem especially promising. It is very important
that all these experiments are mutually complementary, since they may
lead to the values of the Rydberg constant of comparable accuracy based on
the different kinds of experimental data. On the theoretical side
calculation of the still unknown corrections to the energy levels discussed
above with the goal of reduction of the theoretical error in determination
of the $1S$ Lamb shift to the level of 1 kHz  (and, respectively, of the
$2S$ Lamb shift to several tenth of kHz) seems to be both quite perspective
and feasible.

\medskip

We are deeply grateful to S. G. Karshenboim who took part at the initial
stage of this project as described in \cite{eksl1,eksl2}. M. E. is deeply
grateful to H. Grotch for fruitful collaboration on the other corrections of
order $\alpha^2(Z\alpha)^5$ to the Lamb shift and for numerous helpful
discussions. We are indebted to P. Mohr for helpful communications on
the vacuum polarization contributions to the Lamb shift and for providing us
his new results prior to publication. We highly appreciate useful
communications with M. Boshier on the preliminary results of the new
measurement of the $1S$ Lamb shift \cite{bosh}. M. E. acknowledges helpful
discussions on the physics of hyperfine splitting and Lamb shift with T.
Kinoshita, D. Owen and K. Pachucki.

The analytic calculations above were partially performed with the help
of the REDUCE. Final numerical calculations with the help of the VEGAS
program were conducted at the Maui High Performance Computing Center
and we highly appreciate the opportunity to use these facilities.
Through the use of the MHPCC, this research was sponsored in part by the
Phillips Laboratory, Air Force Materiel Command, USAF, under cooperative
agreement number F29601-93-2-0001. The views and conclusions contained
in this document are those of the authors and should not be
interpreted as necessarily representing the official policies or
endorsements, either expressed or implied, of Phillips Laboratory or
the U.S. Government. Part of this work was done during the visit of M.E. to
the Penn State University. He is deeply grateful to the colleagues at the
Physics Department of the Penn State University and especially to H. Grotch
for kind hospitality. The research described in this publication was made
possible in part by grant \#R2E000 from the International Science Foundation
and was also supported by the Russian Foundation for Fundamental Research
under grant \#93-02-3853. Work of M.E. was supported in part by the
National Science Foundation under grant \#NSF-PHY-9120102.

\newpage

\begin{center}
\underline{Table 1. Experimental Results}
\end{center}
\nopagebreak \vspace*{4pt}
\begin{center}

\begin{tabular}{|l|ll|}    \hline
\ $Interval$  & $\Delta E\mbox{ (kHz)}$   &
\\ \hline
$\mbox{Hydrogen, $1S_{1/2}$ \cite{weitz}}$      &   $8~172~860~(60)$    &
\\ \hline
$\mbox{Hydrogen, $1S_{1/2}$ \cite{nez}}$      &   $8~172~815~(70)$   &
\\ \hline
$\mbox{Hydrogen, $1S_{1/2}$ \cite{bosh}}$     &   $8~172~844~(55)~$    &
\\ \hline
$\mbox{Hydrogen, $2S_{1/2}-2P_{1/2}$ \cite{lp}}$   & $1057~845~(9)$     &
\\ \hline
$\mbox{Hydrogen, $2S_{1/2}-2P_{1/2}$ \cite{sok,kar1}}$   &
$1057~857.6~(2.1)$ & \\ \hline
$\mbox{Hydrogen, $2S_{1/2}-2P_{1/2}$ \cite{hp}}$   & $1057~839~(12)$    &
\\ \hline
\ $\mbox{Muonium, HFS \cite{mariam}}$  & $4~463~302.88~(16)$   &
\\ \hline
\end{tabular}
\end{center}

\newpage
\begin{center}
\underline{Table 2. Corrections to HFS}
\end{center}
\begin{center}
\underline{and Lamb shift}
\nopagebreak \vspace*{4pt}

\begin{tabular}{|l|r|r|}    \hline
 $$  &HFS                &        Lamb shift\\
$$  & $\frac{\alpha^2(Z\alpha)}{\pi n^3}E_F$
&$\frac{\alpha^2(Z\alpha)^5}{\pi n^3}(\frac{m_r}{m})^3m$ \\
       &                    &
\\ \hline
$a$    &   $9/4        $    & $0$\\ \hline
$b$    &   $-6.65997(1)$    & $2.9551(1)$\\ \hline
$c$    &   $3.93208(1) $    & $-2.2231(1)$\\ \hline
$d$    &   $-3.903368(79)$  & $-5.238023(56)$\\ \hline
$e$    &   $4.566710(24)$   & $5.056278(81)$\\ \hline
$f$    &   $-3.404163(22)$  & $-1.016145(21)$\\ \hline
$g$    &   $2.684706(26)$   & $-0.1460233(52)$\\ \hline
$h$    &   ${33}/{16}  $    & $153/80$\\ \hline
$i$    &   $0.05524(21)$    & $-5.51680(87)$\\ \hline
$j$    &   $-7.14860(39)$   & $-7.76648(79)$\\ \hline
$k$    &   $1.465834(20)$   & $1.959589(33)$\\ \hline
$l$    &   $-1.983298(95)$  & $1.74815(38)$\\ \hline
$m$    &   $3.16956(16)$    & $1.87541(49)$\\ \hline
$n$    &   $-3.59566(14)$   & $-1.30626(49)$\\ \hline
$o$    &   $1.80491(45)$    & $-12.0641(16)$\\ \hline
$p$    &   $3.50608(16)$    & $6.13527(90)$\\ \hline
$q$    &   $-0.80380(15)$   & $-7.52272(83)$\\ \hline
$r$    &   $1.05298(18)$    & $14.3622(15)$\\ \hline
$s$    &   $0.277203(27)$   & $-0.930291(78)$\\ \hline\hline
$Total$&   $-0.6711(7)$     & $-7.725(3)$\\ \hline
\end{tabular}
\end{center}

\newpage
\begin{center}
\underline{Table 3. Some New Contributions to the Lamb Shift}
\end{center}
\nopagebreak \vspace*{4pt}
\begin{center}

\begin{tabular}{|l|ll|}    \hline
\ $Level$  &    &$\Delta E$
\\ \hline
$nS_{1/2} \cite{kar2}$   &&$-\frac{8}{27}\frac{\alpha^2(Z\alpha)^6}{\pi^2n^3}
\cdot{\ln^3[Z\alpha]^{-2}}(\frac{m_r}{m})^3m$
\\ \hline
$nS_{1/2} \cite{grp}$   &&$(4\ln
2-\frac{7}{2})\frac{(Z\alpha)^6}{n^3}\frac{m}{M}m$ \\ \hline
$nS_{1/2} \cite{eg}$
&&$(\frac{2\pi^2}{9}-\frac{70}{27})
\frac{m}{M}\frac{\alpha(Z\alpha)^5}{\pi^2n^3}(\frac{m_r}{m})^3m$ \\ \hline
$nP_j$ \cite{gkme}&&$\{[(\frac{217}{480} + \frac{3}{16n} - \frac{14}{15n^2}
+ \frac{1}{2n^3} - \sigma{\bf l}(\frac{7}{192} + \frac{3}{32n}$ \\
&&$ -\frac{1}{6n^2})]\frac{(Z\alpha)^6}{n^3} + \sigma{\bf l}\frac{0.328}{3}
\frac{\alpha^2(Z\alpha)^6}{\pi^2n^3}\}\frac{m}{M}m$ \\ \hline
$nS_{1/2}$
&&$-4[\frac{1}{15}(\frac{m}{m_\mu})^2+\Sigma_{v_i}\frac{4\pi^2}{f^2_{v_i}
m^2_{v_i}}+\frac{2}{3}\frac{1}{\mbox{1 GeV}^2}]
\frac{\alpha(Z\alpha)^4}{\pi n^3}m$ \\
\hline
\end{tabular}
\end{center}

\begin{center}
\underline{Table 4. $2S_{1/2}-2P_{1/2}$ Lamb Shift}
\end{center}
\nopagebreak \vspace*{4pt}
\begin{center}

\begin{tabular}{|l|ll|}    \hline
\ \ Source of the value  & $\Delta E\mbox{ (kHz)}$   &
\\ \hline
$\mbox{Experimental result \cite{lp}}$      &   $1~057~845~(9)$    &
\\ \hline
$\mbox{Experimental result \cite{sok,kar1}}$&   $1~057~857.~6~(2.1)$   &
\\ \hline
$\mbox{Experimental result \cite{hp}}$     &   $1~057~839~(12)~$    &
\\ \hline
$\mbox{Theory, $r_{p}=0.805~(11)$~fm~\cite{dhr}}$   & $1057~810~(4)~(4)$ &
\\ \hline
$\mbox{Theory, $r_{p}=0.862~(12)$~fm~\cite{ssbw}}$ & $1057~829~(4)~(4)$ &
\\ \hline
\ $\mbox{Self-consistent 1S \cite{weitz}} $  & $1057~854~(16)$   &
\\ \hline
\ $\mbox{Self-consistent 1S \cite{nez,and}} $  & $1057~835~(15)$   &
\\ \hline
\ $\mbox{Self-consistent 1S \cite{bosh}} $  & $1057~847~(13)$   &
\\ \hline
\end{tabular}
\end{center}

\begin{center}
\underline{Table 5. Self-Consistent 1S Lamb shift values}
\end{center}
\nopagebreak \vspace*{4pt}
\begin{center}

\begin{tabular}{|l|ll|}    \hline
\ Source of the value  & $\Delta E\mbox{ (kHz)}$   &
\\ \hline
$\mbox{Ref. \cite{weitz}}$      &   $8~172~915~(129)$    &
\\ \hline
$\mbox{Ref. \cite{nez}}$      &   $8~172~763~(117)$   &
\\ \hline
$\mbox{Ref. \cite{bosh}}$     &   $8~172~858~(107)$    &
\\ \hline
$\mbox{Theory, this work}$            & $8~172~729~(28)~(32)$     &
\\ \hline
\end{tabular}
\end{center}

\newpage

\begin{center}
\underline{Table 6. Rydberg Constant}
\end{center}
\nopagebreak \vspace*{4pt}
\begin{center}
\begin{tabular}{|l|ll|}    \hline
\ Source of the value & $R_\infty~\mbox{(cm$^{-1}$)}$   &
\\ \hline
$\mbox{Experiment \cite{and}}$&   $109~737.~315~684~1~(42)$   &
\\ \hline
$\mbox{Experiment \cite{nez}}$&   $109~737.~315~683~4~(24)$    &
\\ \hline
$\mbox{Self-consistent \cite{and,weitz}}$&$109~737.~315~686~8~(58)~(20)$ &
\\ \hline
$\mbox{Self-consistent \cite{nez,and}}$ & $109~737.~315~681~1~(52)~(14)$ &
\\ \hline \
$\mbox{Theory and \cite{and}} $  & $109~737.~315~679~7~(12~)(20)~(14)$   &
\\ \hline
\ $\mbox{Theory and \cite{nez}} $ & $109~737.~315~680~2~(05)~(14)~(06)$   &
\\ \hline
\end{tabular}
\end{center}

\newpage
\begin{center}
{Figure Captions}
\end{center}

Fig. 1. Six gauge invariant sets of graphs producing corrections of order
$\alpha^2(Z\alpha)^5$.\\

Fig. 2. Nineteen topologically different diagrams with two radiative photon
insertions in the electron line.

\end{document}